\documentclass[aps,pra,twocolumn,superscriptaddress,nofootinbib]{revtex4-1}
\usepackage{amsmath}
\usepackage{amssymb}
\usepackage{graphicx}
\usepackage{mathrsfs}
\usepackage{ntheorem}
\usepackage[T1]{fontenc}
\usepackage{bm}
\usepackage{subfigure}
\usepackage[bookmarks=false]{hyperref}
\usepackage{color}
\hypersetup{colorlinks=true,citecolor=blue,linkcolor=blue,urlcolor=blue,pdfstartview=FitH,bookmarksopen=true}

\newcommand{\ket}[1]{|#1\rangle}
\newcommand{\bra}[1]{\langle #1|}
\newcommand{\inp}[2]{\langle #1|#2\rangle}
\newcommand{\expectation}[1]{\langle #1 \rangle}
\newcommand{\Tr}{\mathrm{Tr}}
\newcommand{\ud}{\mathrm{d}}

\newcommand{\abs}[1]{\lvert #1\rvert}

\def\CC{{\rm\kern.24em \vrule width.04em height1.46ex depth-.07ex \kern-.30em C}}
\def\RR{{\rm\kern.24em \vrule width.04em height1.46ex depth-.07ex
\kern-.30em R}}
\def\P{{\rm I\kern-.25em P}}

\begin{document}
\title{Topological characterization of one-dimensional open fermionic systems}
\author{Da-Jian Zhang}
\affiliation{Department of Physics, National University of Singapore, Singapore 117542}
\author{Jiangbin Gong}
\email{phygj@nus.edu.sg}
\affiliation{Department of Physics, National University of Singapore, Singapore 117542}

\date{\today}

\begin{abstract}
A topological measure characterizing symmetry-protected topological phases
in one-dimensional open fermionic systems is proposed.
It is built upon the kinematic approach to the geometric phase of mixed states
and facilitates the extension of  the notion of topological phases from zero-temperature
to nonzero-temperature cases. In contrast to a previous finding that topological properties
may not survive above a certain critical temperature,
we find that topological properties of open systems, in the sense of the measure
suggested here, can persist at any
finite temperature and disappear only in the mathematical limit of infinite temperature.
Our result is illustrated with two paradigmatic models of topological
matter. The bulk topology at nonzero temperatures manifested as robust mixed edge
state populations is examined via two figures of merit.
\end{abstract}

\maketitle


Berry phase is a fundamental concept in quantum physics, revealing a
gauge field governing parallel transport (originally for pure states) \cite{1984Berry45}.
It was later realized that the Berry phase of electronic wave functions has
a profound effect on topological
properties of materials and is responsible for a number of
topological phenomena \cite{2010Xiao1959}. A seminal example is the
integer quantum Hall effect, which is related to the Berry phase for
a contour enclosing a two-dimensional Brillouin zone and determines the quantized value
of the Hall conductivity of filled bands  \cite{1982Thouless405}. Besides
the integer quantum Hall effect, several
fascinating discoveries in modern condensed-mater
physics, including topological insulators and topological
superconductors \cite{2010Hasan3045,2011Qi1057}, are also deeply
connected to the Berry phase.

Any realistic quantum system at a nonzero temperature, however, inevitably interacts with its
environment and is described by mixed states rather than pure states.
This historically stimulated the development towards extending the Berry phase
to the realm of mixed states. Uhlmann was the first to address the issue of
mixed-state holonomy. He formulated a new parallel transport condition defining
parallelity and holonomy for mixed states, with which the Uhlmann phase was
put forward \cite{1986Uhlmann229,1991Uhlmann229,1993Uhlmann253}.
Another definition of the geometric phase for mixed states is proposed by
Tong \textit{et al.}, based on a kinematic approach, without a priori assumption
about the dynamics of open systems  \cite{2004Tong80405}. In addition,
there have been other formulations of parallel transport conditions for mixed
states and several alternative but nonequivalent definitions have been proposed
accordingly \cite{2000Sjoeqvist2845,2003Whitney190402,2003Ericsson90405,2003Filipp50403,
2003Carollo160402,2004Carollo20402,2005Whitney70407}.

Armed with the concept of geometric phase for mixed states, it becomes possible to investigate
the role of thermal and dissipative effects in topological phases of matter. In particular, what is the impact of a nonzero temperature on the topological characterization of an otherwise topologically nontrivial system at zero temperature?
This topic has received much attention recently \cite{2011Diehl971,2013Rivas155141,2014Viyuela130401,
2014Viyuela76408,2014Huang76407,2014Nieuwenburg75141,2015Budich165140,2015Viyuela34006,
2017Mera15702,2018Viyuela10,2018He235141,
2018Bardyn11035,2018Caldas17,2018Paunkovic11129,Amin,Leonforte}, with the
Uhlmann phase used as the main tool. Viyuela \textit{et al.} introduced the Uhlmann phase
as a topological measure for one-dimensional (1D) open fermionic systems \cite{2014Viyuela130401}.
For physical models considered there, topological properties in
the sense of their topological measure were found to disappear above
a certain critical temperature, at which a certain Uhlmann phase goes discontinuously and abruptly
to zero. Later on, similar results were found in
two-dimensional  open fermionic systems, with the aid of proper topological invariants
constructed out of the Uhlmann phase \cite{2014Viyuela76408,2014Huang76407}.
Although much progress has been made by use of the Uhlmann phase,
it is just the beginning to study thermal
topological behavior with respect to other legitimate definitions
of mixed-state geometric phase. Indeed, to our best knowledge, attempts were made only for the interferometric phase
\cite{2000Sjoeqvist2845}, and the associated calculations for the Kitaev chain \cite{2016Andersson20150231,2018Bhattacharya} yield results different from those obtained from the Uhlmann phase approach. Despite the lack of extensive studies, the existing results already indicate that the physical picture of thermal topological phases of matter may not be fully
captured by the Uhlmann phase, and therefore, more efforts concerning other definitions of mixed-state geometric phase are  desirable.

In this paper, based on the kinematic approach to the geometric phase
for mixed states \cite{2004Tong80405}, we introduce a topological measure to
characterize symmetry-protected topological phases in 1D open fermionic systems.
It places the temperature under equal footing with other tuneable system parameters,
thus providing a way for extending the notion of topological phases
to cases with nonzero temperature. To reveal the physical implication of our
topological measure, we further introduce two figures of merit, which can measure the degree
of the presence of mixed edge states. As examples,
we use our
topological measure to identify the bulk topological invariants of two paradigmatic models of
symmetry protected topological matter under arbitrary temperature. We also investigate the physical implication
of the bulk topology for robust mixed edge states
with the aid of our figures of merit.
Although the derivation of our topological measure is relatively
simple, it depicts an interesting physical picture of thermal topological
phases different from that in the previous work \cite{2014Viyuela130401}.


To present our finding clearly, we need first
to recapitulate the kinematic approach to the geometric phase for mixed states.
Consider an open quantum system $s$ equipped with a $N$ dimensional Hilbert
space $\mathcal{H}_s$. An evolution of the state of $s$ can be expressed as
the path
\begin{eqnarray}\label{eq:path1}
\mathcal{P}: t\in[0,\tau]\mapsto\rho(t)=\sum_{j=1}^N p_j(t)
\ket{\phi_j(t)}\bra{\phi_j(t)},
\end{eqnarray}
where $p_j(t)$ and $\ket{\phi_j(t)}$ are the eigenvalues and
eigenvectors of the density operator $\rho(t)$, respectively. In order to
associate the path $\mathcal{P}$ with a geometric phase, the mixed
state in the path is lifted to a pure state by the standard purification procedure,
\begin{eqnarray}\label{eq:purification}
\ket{\Phi(t)}=\sum_{j=1}^N\sqrt{p_j(t)}\ket{\phi_j(t)}\otimes\ket{a_j},
\end{eqnarray}
where an auxiliary quantum system $a$ is introduced, and $\ket{a_j}$ denotes a fixed
orthonormal basis of its Hilbert space $\mathcal{H}_a$.
Note that there is redundancy in
the expression (\ref{eq:purification}), because of the gauge degree of freedom in
choosing states $\ket{\phi_j(t)}$. If all the nonzero $p_j(t)$ are
non-degenerate during the evolution, the redundancy
in Eq.~(\ref{eq:purification}) can be removed by imposing the following
parallel transport condition on $\ket{\phi_j(t)}$,
\begin{eqnarray}\label{eq:para-trans}
\bra{\phi_j(t)}\partial_t\ket{\phi_j(t)}=0,~~~~j=1,\dots,N.
\end{eqnarray}
Under the condition of Eq. (\ref{eq:para-trans}), the state $\ket{\Phi(t)}$
experiences the Berry-Simon parallel transport, i.e.,
$\bra{\Phi(t)}\partial_t\ket{\Phi(t)}=0$ \cite{1984Berry45,1983Simon2167}.
The acquired relative phase between $\ket{\Phi(\tau)}$ and $\ket{\Phi(0)}$ is
defined to be the geometric phase associated with the path $\mathcal{P}$, namely, $\gamma[\mathcal{P}]:=\arg\langle\Phi(0)|\Phi(\tau)\rangle$.
Substituting Eq. (\ref{eq:purification}) into this defining expression yields
\begin{eqnarray}\label{df:GP}
\gamma[\mathcal{P}]=\arg\left(\sum_{j=1}^N
\sqrt{p_j(0)p_j(\tau)}\inp{\phi_j^\shortparallel(0)}
{\phi_j^\shortparallel(\tau)}\right),
\end{eqnarray}
with $\ket{\phi_j^\shortparallel(t)}$ satisfying Eq. (\ref{eq:para-trans}). Here, to indicate the fact that the eigenvectors undergo parallel transport, i.e., satisfy Eq. (\ref{eq:para-trans}), we use $\ket{\phi_j^\shortparallel(t)}$ instead of $\ket{\phi_j(t)}$ to represent it.
Moreover, a gauge-invariant expression for $\gamma[\mathcal{P}]$ reads
\begin{widetext}
\begin{eqnarray}\label{eq:geom-phase}
\gamma[\mathcal{P}]=\arg\left(\sum_{j=1}^N\sqrt{p_j(0)p_j(\tau)}
\inp{\phi_j(0)}{\phi_j(\tau)}e^{-\int_0^\tau\bra{\phi_j(t)}\partial_t\ket{\phi_j(t)}\ud t}\right),
\end{eqnarray}
\end{widetext}
where $\ket{\phi_j(t)}$ is not necessary to fulfill Eq. (\ref{eq:para-trans}). The invariance of Eq. (\ref{eq:geom-phase}) under gauge transformations $\ket{\phi_j(t)}\mapsto e^{i\theta_j(t)}\ket{\phi_j(t)}$, $\theta_j(t)$ real, can be checked straightforwardly.
In the case that some of the nonzero $p_j(t)$ are degenerate,
the path $\mathcal{P}$ may be rewritten as
\begin{eqnarray}\label{eq:path2}
\mathcal{P}: t\in[0,\tau]\mapsto\rho(t)=\sum_{j=1}^K\sum_{\mu=1}^{n_j}
p_j(t)\ket{\phi_j^\mu(t)}\bra{\phi_j^\mu(t)},
\end{eqnarray}
where $p_j(t)$, $j=1,\dots,K$, are the eigenvalues of $\rho(t)$ with
degeneracy $n_j$, and $\ket{\phi_j^\mu(t)}$, $\mu=1,\dots,n_j$, are the
corresponding degenerate eigenvectors. In this case, the parallel transport
condition in Eq. (\ref{eq:para-trans}) is no longer capable of completely removing the
redundancy in the standard purification procedure. As a consequence, it should be replaced by
\begin{eqnarray}\label{eq:para-trans2}
\bra{\phi_j^\mu(t)}\partial_t\ket{\phi_j^\nu(t)}=0,~~~~\mu,\nu=1,\dots,n_j,
\end{eqnarray}
while the remainder of the process of defining $\gamma[\mathcal{P}]$ is unchanged.

With the above knowledge, we now move on to our topic of introducing a
topological measure for 1D open fermionic systems.

For simplicity, we consider a two-band single-particle model describing a topological insulator.
Under periodic boundary condition, the Hamiltonian of the model can be generally
expressed as
\begin{eqnarray}
H=\sum_{k}\psi^\dagger(k)H(k)\psi(k),
\end{eqnarray}
where $H(k):=d_0(k)+\sum_{i=1}^3 d_i(k)\sigma_i$ is the bulk momentum-space
Hamiltonian and $\psi(k):=(a_k,b_k)^t$ stands for the spinor representation,
with $\sigma_i$ denoting the Pauli matrices and $a_k$, $b_k$ representing two
species of fermionic annihilation operators. Our discussion below can be easily
generalized to models of superconductors by using the
Nambu spinor \cite{2010Altland} instead.

When a non-vanishing gap exists in the energy spectrum of $H(k)$,
the vector $\bm{d}(k):=(d_1(k),d_2(k),d_3(k))$ is non-zero for all $k$, and
its normalized vector $\hat{\bm{d}}(k):=\bm{d}(k)/\abs{\bm{d}(k)}$ defines a
mapping from the Brillouin zone into the unit sphere $S^2$. This mapping
is, however, topologically trivial, as the first homotopy group of $S^2$ is trivial.
Hence, symmetry is needed, in order to put further constraints
on $\hat{\bm{d}}(k)$ and induce a non-trivial topology. For 1D fermionic systems,
the symmetry needed is typically the chiral symmetry, i.e.,
a unitary matrix $\Gamma$ satisfying $\Gamma^2=1$ and $\Gamma H(k)\Gamma=-H(k)$ \cite{1997Altland1142,2010Ryu65010,2016Chiu35005}.

Under the natural assumption of the system being in thermodynamic equilibrium
with a reservoir, its state is given by the grand canonical ensemble,
$\rho=\frac{1}{Z}e^{-\beta(H-\mu N)}$,
where $Z$ is the partition function, $N$ the total number operator, $\mu$ the
chemical potential, and $\beta:=\frac{1}{k_bT}$ the reciprocal of the
temperature $T$, with $k_b$ being the Boltzmann's constant. Here we have assumed that there are no interactions
when two or more particles are present in the system.  Note also that it is possible (but not always)
for the grand canonical ensemble to emerge as the unique steady state from the
Lindbladian dynamics \cite{2013Rivas155141}.
In the single-particle picture, the state of the system is described by
the one-body density matrix \cite{2004Cheong75111}
\begin{eqnarray}\label{eq:state}
\rho(k)=\sum_{j=\pm}\omega_j(k)\ket{\mu_j(k)}\bra{\mu_j(k)},
\end{eqnarray}
where $\ket{\mu_j(k)}$, $j=\pm$, are the eigenvectors
of $H(k)$, i.e., $H(k)\ket{\mu_j(k)}=\varepsilon_j(k)\ket{\mu_j(k)}$,
and $\omega_j(k)$ are the Fermi weights expressed as $\omega_j(k)=1/[e^{\beta(\varepsilon_j(k)-\mu)}+1]$.

To arrive at our topological measure, we need to discuss the geometric phase
associated with the path
\begin{eqnarray}
\mathcal{P}: k\in[-\pi,\pi]\mapsto\rho(k)
\end{eqnarray}
case by case. Here, in the spirit of Zak's phase \cite{1989Zak2747},
the Bloch quasimomentum $k$ plays the role of $t$ appearing in Eq. (\ref{eq:path1}).
Consider first the case of arbitrary $\beta>0$ and $\varepsilon_-(k)\neq\varepsilon_+(k)$
for all $k$, i.e., any finite temperature and non-vanishing bulk gap. Adopting a cyclic gauge, i.e.,
$\ket{\mu_j(-\pi)}=\ket{\mu_j(\pi)}$, we have that the Berry phase associated
with $\ket{\mu_j(k)}$ reads $\gamma_j=i\int_{-\pi}^\pi\ud k\bra{\mu_j(k)}\partial_k\ket{\mu_j(k)}$.
Substituting $\omega_j(k)$ and $\ket{\mu_j(k)}$ into Eq. (\ref{eq:geom-phase}) and using the equalities $\omega_j(-\pi)=\omega_j(\pi)$ and $\ket{\mu_j(-\pi)}=\ket{\mu_j(\pi)}$, we have
\begin{eqnarray}\label{eq:step}
\gamma[\mathcal{P}]=\arg[\omega_{-}(\pi)e^{i\gamma_{-}}+
\omega_+(\pi)e^{i\gamma_+}].
\end{eqnarray}
On the other hand, by representing $\Gamma$ in the form $\Gamma=\bm{n}\cdot\bm{\sigma}$
with $\bm{n}$ a constant vector of unit length, the chiral symmetry
condition $\Gamma H(k)\Gamma=-H(k)$ gives $\bm{n}\cdot\bm{d}(k)=0$.
It follows that $\bm{d}(k)$ is restricted to the plane perpendicular to $\bm{n}$, and $\hat{\bm{d}}(k)$ now defines a mapping from the Brillouin zone
into the circle $S^1$. The topology of such a mapping is characterized by
the winding number,
$\nu=\frac{1}{2\pi}\int_{-\pi}^\pi\ud k[\hat{\bm{d}}(k)\times\partial_k
\hat{\bm{d}}(k)]\cdot\bm{n}$, counting the number of times
that $\hat{\bm{d}}(k)$ travels around the origin as $k$ goes through
the Brillouin zone.
It is not difficult to see that the relationship between the Berry phase and
the winding number is $\gamma_-=-\gamma_+=\pi\nu$ up to an integer multiple of $2\pi$.
Substituting this equality into Eq. (\ref{eq:step}) and simplifying the resultant equation by using $\omega_j(\pi)>0$, we obtain
\begin{equation}
\gamma[\mathcal{P}]=\arg[\cos(\pi\nu)].
\end{equation}

Consider next the case of $\beta=0$, i.e., infinite temperature, which is mathematically allowed but not so physical. Here, the spectrum $\varepsilon_j(k)$ may be gapped or gapless. In this case, all the $\omega_j(k)$ in Eq.~(\ref{eq:state}) equals $1/2$ and hence $\mathcal{P}: k\in[-\pi,\pi]\mapsto\rho(k)$ represents a path with degeneracy.
Therefore, we need to find the eigenvectors $\ket{\mu_j(k)}$ of $\rho(k)$ that
satisfy Eq. (\ref{eq:para-trans2}). Without loss of generality, $\ket{\mu_j(k)}$
can be chosen as $\ket{\mu_-(k)}=(1,0)^t$ and $\ket{\mu_+(k)}=(0,1)^t$.
Inserting the two expressions into Eq. (\ref{df:GP}) yields
\begin{eqnarray}
\gamma[\mathcal{P}]=0.
\end{eqnarray}

Finally, for the last case of $\beta>0$ and $\varepsilon_-(k)=\varepsilon_+(k)$
for some $k$, i.e., an arbitrary finite temperature but with a gapless spectrum, there may exist gauges $\ket{\mu_j(k)}$
such that $\inp{\mu_j(-\pi)}{\mu_j(\pi)}=0$, $j=\pm$ \cite{note1}. From Eq. (\ref{eq:geom-phase}), it follows immediately that $\gamma[\mathcal{P}]=\arg 0$, indicating that
$\gamma[\mathcal{P}]$ is ill-defined
in general. Now, combining the three cases, we obtain the topological measure:
\begin{eqnarray}\label{eq:topo-measure}
\gamma[\mathcal{P}]=(1-\delta_{\beta,0})\arg[\cos(\pi\nu)],
\end{eqnarray}
which is defined for the case of arbitrary finite temperature but with gapped spectrum or for the case of arbitrary spectrum but with the infinite temperature.

It is worth noting that several conditions have been speculated to be necessary for a functional to be a legitimate
topological measure in the setting considered here
\cite{2013Rivas155141,2014Viyuela130401,2014Huang76407}.
(i) In the limit of zero
temperature, it should reduce to usual topological order parameters, e.g., the
Berry phase. (ii) It should not increase in the course of increasing
temperature, i.e., topological order cannot be created by temperature. (iii)
In the limit of infinite temperature, it should be zero, i.e., topological order
must be spoiled completely. It can be verified our measure proposed in
Eq.~(\ref{eq:topo-measure}) fulfills all the conditions.  {Besides, it is also worth comparing the derivation presented here with that in Ref. \cite{2014Viyuela130401}. Both the derivation here and that in Ref. \cite{2014Viyuela130401} are on the basis of purifications of mixed states. The purification in Ref. \cite{2014Viyuela130401} is to express $\rho(t)$ as
$\rho(t)=w(t)w^\dagger(t)$, with $w(t)$ being an operator. In contrast, the procedure involved here is given by Eq. (\ref{eq:purification}), which aims to express $\rho(t)$ as a superposition of pure states via an auxiliary quantum system.   
Therefore, by construction,  our topological measure is distinct from the method proposed in Ref. \cite{2014Viyuela130401}.
As far as we can see, there is no reason to prefer one kind of purification of mixed states but abandon the other, which is one of the motivations of this paper.}

Our topological measure is very simple.
For any finite temperature, if $\nu=0$ then $\gamma[\mathcal{P}]=0$; and if $\nu=1$ then $\gamma[\mathcal{P}]=\pi$. Only for the limit of infinite temperature (i.e., $\beta=0$), $\gamma[\mathcal{P}]=0$, regardless of whether the zero-temperature phase is topologically trivial or nontrivial. Despite the simplicity of our measure, it produces some interesting implications for the population of
mixed edge states. Before showing this, let us first elaborate on the topological characterization of two
paradigmatic models using our measure.

\textit{Example 1: SSH model.}---The Su-Schrieffer-Heeger (SSH) model describes
spin-polarized electrons hopping on a dimerized chain \cite{1979Su1698,1982Rice1455}.
In the language of  Altland-Zirnbauer classification \cite{1997Altland1142,2010Ryu65010,2016Chiu35005},
it belongs to the BDI class. The Hamiltonian of the model reads
\begin{eqnarray}\label{H:SSH model}
H_{\textrm{SSH}}=\sum_{j=1}^L (J+\alpha)a_j^\dagger b_j+(J-\alpha)a_j^\dagger b_{j-1}+\textrm{H.c.},
\end{eqnarray}
where $a_j$ and $b_j$ are the fermionic annihilation operators acting on
the $j$-th unit cell which hosts two sites, one on sublattice $A$
and the other on sublattice $B$.
$J+\alpha\geq 0$ and $J-\alpha\geq 0$ denote the intracell
and intercell hopping amplitudes, respectively, with $\alpha$ characterizing
the imbalance between them. Hereafter, we set $J=1$. It is known that $\nu=1$ for
$\alpha<0$ and $\nu=0$ for $\alpha>0$. From this, it follows that $\gamma[\mathcal{P}]=\pi$
for $\alpha<0$ and $\beta>0$, corresponding to the topologically non-trivial region,
and $\gamma[\mathcal{P}]=0$ for $\alpha>0$ and $\beta>0$ or $\beta=0$, corresponding
to the topologically trivial region.

\textit{Example 2: Creutz ladder.}---The Creutz ladder (CL) model describes spin-polarized electrons
moving in a ladder system \cite{1999Creutz2636,2009Bermudez135702}. It belongs to
the AIII class. The Hamiltonian of the model reads
\begin{eqnarray}\label{H:CL model}
H_{\textrm{CL}}=&-&\sum_{j=1}^L[K(e^{-i\theta}a_{j+1}^\dagger a_j+e^{i\theta}b_{j+1}^\dagger b_j)
\nonumber\\
&+&K(b_{j+1}^\dagger a_j+a_{j+1}^\dagger b_j)+M a_j^\dagger b_j+\textrm{H.c.}],
\end{eqnarray}
where $a_j$ and $b_j$ are the fermionic annihilation operators acting on
the $j$-th sites of the upper and lower chain, respectively. The hopping along horizontal and diagonal
links is described by the parameter $K$, and the vertical one by $M$. Additionally,
a magnetic flux $\theta\in[-\pi/2,\pi/2]$ is induced by a perpendicular magnetic field.
Hereafter, we set $K=1$, $\theta=\pi/2$, and redefine a new parameter
$\alpha:=-1+\frac{M}{2K}$. It is known that $\nu=1$ for
$\alpha<0$ and $\nu=0$ for $\alpha>0$. From this, it follows that $\gamma[\mathcal{P}]=\pi$
for $\alpha<0$ and $\beta>0$, corresponding to the topologically non-trivial region,
and $\gamma[\mathcal{P}]=0$ for $\alpha>0$ and $\beta>0$ or $\beta=0$, corresponding
to the topologically trivial region.

A remarkable manifestation of nontrivial bulk topology is the emergence of edge states. For zero-temperature cases
this is known as the bulk-boundary correspondence. Can this physical picture carry over to cases with non-zero temperature?
After introducing our topological measure, it is natural to examine the equilibrium profiles of edge states associated with
topologically distinct regions identified by the measure.

To answer the above question, we need a figure
of merit to characterize the distinguishability
between mixed edge states and bulk states.
Such a figure of merit should be a function of the system parameters and the temperature. For the models considered
throughout this paper, one key system parameter
is $\alpha$ for both models albeit carrying different meanings, and for this reason we shall denote the figure
of merit as $\Lambda(\alpha,\beta)$. Let us assume that $\Lambda(\alpha,\beta)$ is defined as
\begin{eqnarray}\label{eq:fig-merit}
\Lambda(\alpha,\beta):=\abs{\expectation{\overline{N}_{\textrm{edge}}}_\rho-
\expectation{\overline{N}_{\textrm{bulk}}}_\rho},
\end{eqnarray}
leaving the discussion on alternative definition for $\Lambda(\alpha,\beta)$ to the end of this paper.
Here, $\expectation{O}_\rho=\Tr(O\rho)$ stands for expectation values
of observables $O$, and $\rho$ is again the grand canonical ensemble, i.e.,
$\rho=\frac{1}{Z}e^{-\beta(H-\mu N)}$, with $H$ being any one of the
Hamiltonians in Eqs. (\ref{H:SSH model}) and (\ref{H:CL model})
under open boundary conditions. The two observables in Eq. (\ref{eq:fig-merit})
are defined as
$\overline{N}_{\textrm{edge}}:=\sum_{i\in I_e}N_i/\abs{I_e}$ and
$\overline{N}_{\textrm{bulk}}:=\sum_{i\in I_b}N_i/\abs{I_b}$,
where $N_i$ is the particle number operator, and $I_e$ and $I_b$ denote the
index sets for the edge and bulk, respectively. For the SSH model,
$I_e$ is comprised of the site of the first unit cell on sublattice
$A$ and that of the last unit cell on sublattice $B$, while for the
CL model, $I_e$ is comprised of the first sites of the upper and
lower chain and the last sites of the upper and lower chain. $I_b$
is the complementary set of $I_e$. The two expectation values
$\expectation{\overline{N}_{\textrm{edge}}}_\rho$
and $\expectation{\overline{N}_{\textrm{bulk}}}_\rho$ by their very nature
represent the average particle occupation numbers in the
edge and bulk, respectively. Roughly speaking, a non-negligible value of the
figure of merit implies that the average particle occupation number in the edge
is distinguishable from that in the bulk.
This indicates the presence of mixed edge states. By contrast,  a negligible value of our figure of merit
represents that the average particle occupation number in the edge
is barely distinguishable from that in the bulk, thus
indicating the absence of mixed edge states.

With the figure of merit in Eq. (\ref{eq:fig-merit}), we
now study the profiles of edge states of the two models,
with respect to different topological regions identified by our measure.
Henceforth, we set the chemical potential $\mu=0.1$ without loss of generality.

Consider first the SSH model. For the flat-band case, i.e.,
$\alpha=\pm 1$, we analytically calculate the figure of merit in
Eq. (\ref{eq:fig-merit}). After some algebra, it is found that
\begin{eqnarray}\label{eq:ana-cal1}
\Lambda(-1,\beta)=\frac{1}{e^{-\beta\mu}+1}-
&&\frac{1}{2}\left[\frac{1}{e^{-\beta(\mu-2)}+1}+
\frac{1}{e^{-\beta(\mu+2)+1}}\right],\nonumber\\
\end{eqnarray}
with $(\alpha,\beta)=(-1,\beta)$
belonging to the topologically non-trivial region
for $\beta>0$ and the topologically trivial region for $\beta=0$, and
\begin{eqnarray}\label{eq:ana-cal2}
\Lambda(1,\beta)=0,
\end{eqnarray}
with $(\alpha,\beta)=(1,\beta)$ belonging to the topologically trivial region for all $\beta\geq 0$.
Note that $\Lambda(\alpha,\beta)>0$ indicates the presence of edge
states while $\Lambda(\alpha,\beta)=0$ implies the absence of edge states,
and also note that $\Lambda(-1,\beta)>0$ when $\beta>0$ and
$\Lambda(-1,\beta)=0$ when $\beta=0$. From Eqs. (\ref{eq:ana-cal1})
and (\ref{eq:ana-cal2}), it follows immediately that
edge states are present in the topologically non-trivial region while absent in
the topologically trivial region.

\begin{figure}[htbp]
\includegraphics[width=0.3\textwidth]{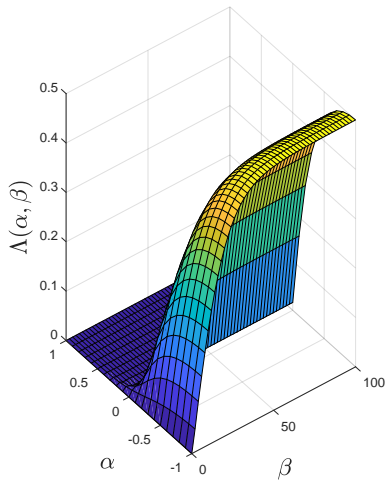}
\caption{Figure of merit for the SSH model. Parameters used are: $L=100$,
$J=1$, and $\mu=0.1$.}
\label{fig:SSH}
\end{figure}

For the general case, analytical solutions may be difficult to obtain and we resort
to numerical computation. Figure \ref{fig:SSH} shows
$\Lambda(\alpha,\beta)$ for the
parameters given in the figure caption.
According to whether the value of
$\Lambda(\alpha,\beta)$ is negligible or not, we can divide the
domain of $\Lambda(\alpha,\beta)$
into two regions. As can be seen in Fig. \ref{fig:SSH}, the region associated with negligible values of $\Lambda(\alpha,\beta)$, referred to as the negligible region for convenience, is  $\alpha>0$ and $\beta>0$
or $\beta=0$, and the region associated with non-negligible values of $\Lambda(\alpha,\beta)$, referred to as the non-negligible region,
is $\alpha<0$ and $\beta>0$. The negligible and non-negligible regions are identical with the topologically trivial and
non-trivial regions identified by our measure, respectively. Note that non-negligible
(negligible) values of the figure of merit indicate the presence (absence)
of edge states. It implies that edge states are present in the topologically non-trivial region while absent in
the topologically trivial region for the SSH model.  The results here strongly support our use of the figure of merit and the physical relevance of our topological characterization.

\begin{figure}[htbp]
\includegraphics[width=0.3\textwidth]{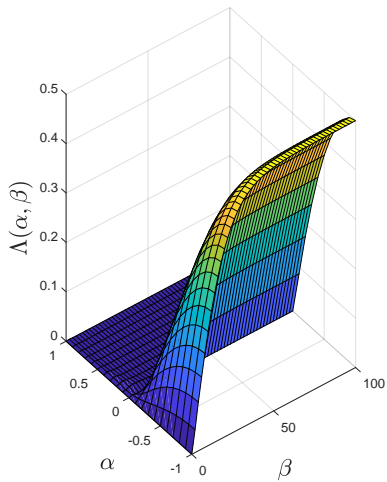}
\caption{Figure of merit for the CL model. Parameters used are:
$L=100$, $K=1$, and $\mu=0.1$.}
\label{fig:CL}
\end{figure}

Consider now the CL model. For the flat-band case, we obtain the same analytical
result as that of the SSH model. For the general case, we resort to numerical
computation. The numerical result is shown in Fig. \ref{fig:CL}, for the
parameters given in the figure caption. Again, we divide the domain of $\Lambda(\alpha,\beta)$ into two regions, according to
whether the value of
$\Lambda(\alpha,\beta)$ is negligible or not. From Fig. \ref{fig:CL}, we deduce that the negligible region is $\alpha>0$ and $\beta>0$ or
$\beta=0$, being identical with the topologically trivial region, and the non-negligible region is
$\alpha<0$ and $\beta>0$, being identical with the topologically non-trivial
region. It indicates that edge states are present in the
topologically non-trivial region while absent in
the topologically trivial region for the CL model, too.

Besides, we have numerically examined the robustness of edge states by adding disorder to the
Hamiltonians in Eqs. (\ref{H:SSH model}) and (\ref{H:CL model}). The numerical results
show that the presence of edge states is insensitive to moderate disorder.

Before concluding, we discuss briefly some aspects of alternative definition of the figure of merit. In the foregoing paragraphs, we have adopted, perhaps, the simplest definition for $\Lambda(\alpha,\beta)$. To characterize the distinguishability between mixed edge and bulk states, one may alternatively adopt the following definition:
\begin{eqnarray}\label{eq:alternative_def}
\Lambda(\alpha,\beta):=\min_{i\in I_e; j\in I_b}\abs{\expectation{N_i}_\rho-\expectation{N_j}_\rho},
\end{eqnarray}
representing the distance between the occupation numbers of electrons in the edge and those in the bulk. It can be shown that the negligible and non-negligible regions with Eq.~(\ref{eq:alternative_def}) are in agreement with the corresponding ones with Eq.~(\ref{eq:fig-merit}), respectively (for the details, see Appendix).
As an immediate consequence, our statements on the bulk-boundary correspondence in the previous paragraphs hold true for the alternative definition in Eq.~(\ref{eq:alternative_def}), too.

In conclusion, we have suggested an alternative and simple measure for characterizing symmetry-protected
topological phases in 1D
open chiral fermionic systems. It
places the temperature under equal footing with other system parameters, thus extending
 the notion of topological phases from zero-temperature to
nonzero-temperature cases. As examples of its application, we have
used our method to identify the bulk topology
of two paradigmatic models of topological matter.

In revealing the physical implication of our simple topological characterization,
we have introduced two figures of merit, which
indicate the presence of mixed edge states when its value is non-negligible.
With them we have shown that mixed edge states are present in the topologically non-trivial
region while absent in the topologically trivial region for the
two models.  We hence conclude that bulk-edge correspondence does hold under our measure, at arbitrary finite temperatures.

Interestingly, in contrast to the
Uhlmann construction \cite{2014Viyuela130401}, for which topological properties cannot
survive above a certain critical temperature, we find that topological properties
in the sense of our measure can persist at any finite
temperature but disappear in the
limit of infinite temperature. Moreover, with our measure the bulk-edge correspondence still holds, whereas it does not exist using the topological invariants under the Uhlmann construction. Our findings are in agreement with the observation made for quasi-local dissipative dynamics in Refs. \cite{2011Diehl971,2013Bardyn85001}, where a topological invariant was formulated in terms of chirally symmetric density matrices and a dissipative bulk-edge correspondence was established. Finally, we would like to point out that
our measure may be experimentally measured with a NMR
quantum simulator \cite{2010Cucchietti240406}.

\begin{acknowledgments}
D.-J. Z. would like to thank Longwen Zhou and Linhu Li for helpful discussions.
This work is supported by Singapore Ministry of Education Academic
Research Fund Tier I (WBS No. R-144-000-353-112) and by the
Singapore NRF grant No. NRFNRFI2017-04 (WBS No. R-144-000-378-281).
D.-J. Z. also acknowledges support from the National Natural Science Foundation of
China through Grant No. 11705105 before he joins NUS.
\end{acknowledgments}

\appendix
\setcounter{equation}{0}

\section*{appendix}\label{app}

\begin{figure}[htbp]
\includegraphics[width=0.3\textwidth]{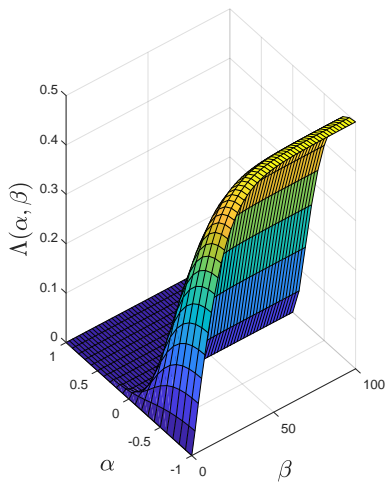}
\caption{Figure of merit for the SSH model with the alternative expression in Eq. (\ref{eq:alternative_def}). Parameters used are: $L=100$,
$J=1$, and $\mu=0.1$.}
\label{fig:SSH-alternative}
\end{figure}

\begin{figure}[htbp]
\includegraphics[width=0.3\textwidth]{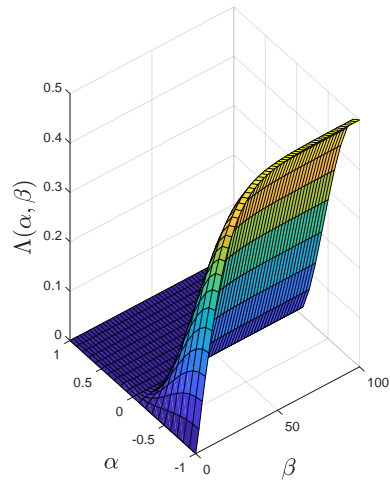}
\caption{Figure of merit for the CL model with the alternative expression in Eq. (\ref{eq:alternative_def}). Parameters used are:
$L=100$, $K=1$, and $\mu=0.1$.}
\label{fig:CL-alternative}
\end{figure}

In this Appendix, we identify the negligible and non-negligible regions of the SSH model and the CL model, with the aid of the alternative definition in Eq. (\ref{eq:alternative_def}).
For the SSH model,
the numerical result is shown in Fig. \ref{fig:SSH-alternative}.
As can be seen in Fig. \ref{fig:SSH-alternative}, the negligible region is $\alpha>0$ and $\beta>0$
or $\beta=0$, and the non- negligible region
is $\alpha<0$ and $\beta>0$.
For the CL model,
the numerical result is shown in Fig. \ref{fig:CL-alternative}.
As can be seen in Fig. \ref{fig:CL-alternative}, the negligible region is $\alpha>0$ and $\beta>0$
or $\beta=0$, and the non-negligible region
is $\alpha<0$ and $\beta>0$.

\clearpage

\begin{thebibliography}{48}
\bibitem [{not()}]{note1}%
  \BibitemOpen
  \href@noop {} {}\bibinfo {note} {To illustrate this point, consider the
  following Hamiltonian \begin{eqnarray}\nonumber H(k)=\begin{pmatrix} 0 &
  1+e^{-ik} \\ 1+e^{ik} & 0 \end{pmatrix}, \end{eqnarray} with gap closing
  points at $k=\pm\pi$. Its eigenvectors read \begin{eqnarray}\nonumber
  \ket{\mu_-(k)}=\frac{1}{\sqrt{2}}\begin{pmatrix} -e^{-ik/2} \\ 1
  \end{pmatrix} , \ket{\mu_+(k)}=\frac{1}{\sqrt{2}}\begin{pmatrix} e^{-ik/2} \\
  1 \end{pmatrix} . \end{eqnarray} Direct calculations show that
  $\inp{\mu_j(-\pi)}{\mu_j(\pi)}=0$ for $j=\pm$.}\BibitemShut {Stop}%
\end{thebibliography}
%

\end{document}